# High-pressure structural behaviour of HoVO$_4$: Combined XRD experiments and *ab initio* calculations

Alka B. Garg[1], D. Errandonea[2,*], P. Rodríguez-Hernández[3], S. López-Moreno[4], A. Muñoz[3], and C. Popescu[4]

[1]*High Pressure and Synchrotron Radiation Physics Division, Bhabha Atomic Research Centre, Mumbai 400085, Maharashtra, India*
[2]*Departamento de Física Aplicada-ICMUV, MALTA Consolider Team, Universidad de Valencia, Edificio de Investigación, C/Dr. Moliner 50, Burjassot, 46100 Valencia, Spain*
[3]*Departamento de Física , Instituto de Materiales y Nanotecnología, and MALTA Consolider Team, Universidad de La Laguna, La Laguna, E-38205 Tenerife, Spain*
[4]*Escuela Superior Cd. Sahagun, Universidad Autónoma del Estado de Hidalgo, Carretera Cd. Sahagun-Otumba s/n 43990, Hidalgo, México*
[5]*CELLS-ALBA Synchrotron Light Facility, Cerdanyola del Valles, 08290 Barcelona, Spain*

* Corresponding author: daniel.errandonea@uv.es

**Abstract.** We report a high-pressure experimental and theoretical investigation of the structural properties of zircon-type HoVO$_4$. Angle-dispersive x-ray diffraction measurements were carried out under quasi-hydrostatic and partial non-hydrostatic conditions up to 28 and 23.7 GPa, respectively. In the first case, an irreversible phase transition is found at 8.2 GPa. In the second case, the onset of the transition is detected at 4.5 GPa, a second (reversible) transition is found at 20.4 GPa, and a partial decomposition of HoVO$_4$ was observed. The structures of the different phases have been assigned and their equations of state (EOS) determined. Experimental results have also been compared to theoretical calculations which fully agree with quasi-hydrostatic experiments. Theory also suggests the possibility of another phase transition at 32 GPa; i.e. beyond the pressure limit covered by present experiments. Furthermore, calculations show that deviatoric stresses could trigger the transition found at 20.4 GPa under non-hydrostatic conditions. The reliability of the present experimental and theoretical results is supported by the consistency between the values yielded for transition pressures and EOS parameters by the two methods.

Keywords: zircon, scheelite, x-ray diffraction, high pressure, phase transition, equation of state

PACS numbers: 62.50.-p, 64.70.K-, 81.30.Hd, 64.10.+h

# 1. Introduction

Holmium orthovanadate, $HoVO_4$, belongs to the family of zircon-type orthovanadates (space group: $I4_1/amd$, Z = 4). Under compression it undergoes a phase transition below 10 GPa [1]. Though the high-pressure phase is crystalline in nature, it's structure is not yet determined accurately. Zircon-type orthovanadates have important applications [2], which include uses in cathodoluminescence and lithium ion batteries, as well as thermophosphors, scintillators, photocatalysis materials, and laser-host materials [3]. They can also be useful for the development of green technologies through applications like photocatalytic hydrogen production [4]. Finally, rare-earth vanadates also show interesting structural and magnetic transformations at low temperatures [5] and in a few of them an incomparable large Jahn-Teller distortion has been encountered [6].

During the last decade, intensive investigations have been carried out on the structural evolution of orthovanadates under high-pressure (HP). They showed that compression is an efficient tool to improve the understanding of the main physical properties of vanadates. In particular, x-ray diffraction [7 - 12], optical [13, 14], and Raman scattering measurements [14 - 18] as well as theoretical calculations [8, 15, 19, 20] have been carried out to understand the structural modifications induced by pressure. Among zircon-type rare-earth orthovanadates, $HoVO_4$ is one of the less studied systems under HP. In particular, it has only been studied by Raman spectroscopy in diamond-anvil cell (DAC) experiments and by *ab initio* calculations. These studies reported a phase transition at 6.1 - 9.3 GPa and proposed that the HP phase has a scheelite-type (space group: $I4_1/a$, Z = 4) structure [1]. This structure appears to be the same as for the sample obtained by quenching from pressures exceeding 3 GPa using uniaxial pressure devices by Stubican and Roy [21]. These results call for the performance of *in-situ* HP x-ray diffraction (XRD) experiments to accurately determine the structure of the HP phase of $HoVO_4$. *Ab initio* calculations can also be helpful for this purpose and for predicting HP phases not discovered yet.

To shed more light on the understanding of the structural properties of zircon-type orthovanadates, and in particular their pressure-induced phase transitions, we have studied the HP behaviour of $HoVO_4$ by XRD up to 28 GPa. *Ab initio* calculations have also been performed up to 50 GPa. Under quasi-hydrostatic conditions we confirmed the zircon-to-scheelite transition, which is found at 8.2 GPa. The structure of the scheelite phase has been Rietveld refined. We also observed that in the presence of deviatoric stresses a second transition took place at 20.4 GPa, being the new HP phase assigned to a monoclinic fergusonite-type structure (space group: $I2/a$, Z = 4). Under these conditions a partial decomposition of $HoVO_4$ is also detected. *Ab initio* calculations fully agree with the experiments and additionally predict the occurrence of a third phase transition at 32 GPa to an orthorhombic structure (space group:

*Cmca*, Z = 8) under hydrostatic conditions. The obtained results are compared with the HP structural behaviour of related orthovanadates. The axial compressibility and room temperature P-V equation of state (EOS) of the different phases of $HoVO_4$ are also reported.

## 2. Experimental Details

Polycrystalline $HoVO_4$ used in the experiments was prepared by solid-state reaction of appropriate amounts of pre-dried $Ho_2O_3$ (Alfa-Aesar 99.9 %) and $V_2O_5$ (Aldrich 99.6 %). The sample obtained was characterized by powder XRD using a Panalytical X-pert Pro diffractometer with Cu $K_\alpha$ radiation. A single phase with the zircon-type structure was confirmed with unit-cell parameters $a$ = 7.123(7) Å and c = 6.289(6) Å, which agrees with those reported in the literature [22]. Two series of experiments were performed at room temperature upon compression: run 1 up to 23.7 GPa using 4:1 methanol-ethanol (ME) as pressure-transmitting medium (PTM) and run 2 up to 28 GPa using Ar as PTM. Experimental conditions can be considered quasi-hydrostatic in run 2 and less hydrostatic in run 1 [23, 24]. Angle-dispersive x-ray diffraction (ADXRD) experiments were carried out using DACs with diamond culets of 300 – 400 μm. The pressure chamber was a 100 – 150 μm hole drilled on hardened stainless steel or inconel gaskets pre-indented to 40 – 60 μm thickness. Powdered $HoVO_4$ samples were loaded in the pressure chamber together with a few Pt grains. The EOS of Pt was used as pressure scale [25]. In run 2, at pressures equal or higher than 7.1 GPa, also the EOS of Ar [23] was used to confirm the pressure obtained from Pt. Pressure was determined with an accuracy of 0.05 GPa. *In-situ* HP XRD measurements were carried out at the XRD1 beam-line of Elettra synchrotron and the MSPD beam-line of ALBA synchrotron [26]. At Elettra, monochromatic x-rays of wavelength 0.6888 Å were used being the beam limited to 80 μm in diameter using a circular collimator. Images of the powder diffraction rings were collected with a MAR345 image-plate. Exposure times of 15–20 minutes were employed for measurements. At ALBA the incident monochromatic beam of wavelength 0.4246 Å was focused down to a 10 μm x 15 μm spot using Kirkpatrick-Baez mirrors. A Rayonix CCD detector was used to collect XRD patterns. In this experiment, thanks to the high-brilliance photon beam of ALBA, exposure time was 10 – 30 seconds. The two dimensional diffraction images collected in both runs were integrated with FIT2D software [27]. Structural analysis was performed with PowderCell [28] and GSAS [29].

## 3. Theoretical Details

First-principles simulations were performed within the framework of Density Functional Theory (DFT) [30] using the Vienna *ab initio* simulations package (VASP) [31, 32]. This package allows obtaining the *ab initio* total energy by means of the plane-wave pseudo-potential method, and computing important properties of the system. Not only the total energy, but also derivatives of the energy, like forces or stresses can be obtained. We use the Projector Augmented Wave (PAW) pseudo-potential scheme [33, 34] that takes into account the full nodal character of the all electron charge density inside the core region. In order to obtain highly converged and accurate results the set of plane-waves employed in our simulations are extended up to a cutoff of 520 eV. It is known that standard DFT cannot work properly for strongly localized *f* electrons. To deal with this problem we used the standard procedure that allows treating *f* electrons using PAW pseudo-potentials [34, 35]. For the Ho atom all the *f* electrons except one are frozen into the core region during the pseudo-potential generation. In our first principles calculations we have described the exchange-correlation energy using the Generalized Gradient Approximation (GGA) with the Perdew-Burke-Ernserhoff (PBE) prescription [36]. Typical underestimation of the cohesion energy coming from GGA approach is well known. It produces an overestimation of the equilibrium volume of the system under study [37]. We used dense grids of k-special points, appropriate to each structure considered in our study, to sample the Brillouin Zone (BZ) to ensure a high convergence of 1-2 meV per formula unit in the total energy of each structure considered in our study. For each selected volume we fully relaxed every structure to their equilibrium configurations through the calculations of the forces on the atoms and the stress tensor [37], until the forces were smaller than 0.006 eV/Å, and the deviation of the stress tensor from a diagonal hydrostatic form was less than 0.1 GPa. This means that our *ab initio* calculations can provide a set of accurate energy, volume and pressure (E, V, P) data that can be fitted using an equation of state (EOS) in order to obtain the equilibrium volume ($V_0$), bulk modulus ($B_0$), and its pressure derivatives ($B_0'$ and $B_0''$). Although our theoretical study was performed under hydrostatic conditions, we also simulated non-hydrostatic conditions for the fergusonite structure. Under hydrostatic conditions the fergusonite structure once relaxed reduces systematically to the scheelite structure. We imposed the non-hydrostatic conditions to the system when we performed the total energy calculations. This allows estimating the possibility of the appearance of the fergusonite structure under pressure when using a non-hydrostatic PTM.

### 4. Results and Discussion

*4.1. High-pressure XRD experiments*

Fig.1 shows XRD patterns measured in run 1 using ME as PTM. In the figure, four Bragg peaks associated to Pt can be easily identified since these peaks have a different pressure evolution than those of the sample. In this run, we found that the patterns obtained from ambient pressure up to 3.2 GPa can be unequivocally assigned to the zircon structure. This is illustrated in the figure by the diffraction pattern measured at 0.6 GPa, for which the residuals of the structural refinement are shown in Fig. 2. The R-factors of the refinement are $R_p$ = 6.88% and $R_{wp}$ = 9.26%. The unit-cell parameters determined at 0.6 GPa are *a* = 7.1074(10) Å and *c* =

6.3076(14) Å. When pressure reaches 4.5 GPa a new Bragg peak emerges near 2θ = 13º. This peak is identified by the symbol $ in Fig. 1a. Upon compression, this and other extra peaks gradually grow in intensity and simultaneously the peaks indexed as zircon $HoVO_4$ gradually vanish. These changes can be assigned to the onset of a phase transition, coexisting the zircon and the HP phase from 4.5 to 13.3 GPa. The HP phase appears as a single phase in a further compression step at 15.5 GPa (see Fig. 1b). The onset pressure of the transition is close to that previously found in Raman experiments carried out using the same PTM (onset 6.2 GPa and pure HP phase at 14.3 GPa) [1]. Fig.1b shows XRD patterns measured upon compression from 15.5 to 23.1 GPa. We found that the HP phase can be assigned to the scheelite structure, as previously proposed from Raman experiments [1]. In this regard, $HoVO_4$ appears to behave in a similar way as other orthovanadates with small trivalent cations; e.g. $LuVO_4$ [12]. The assignment of the scheelite-type structure to the HP phase is supported by structural refinements. The residuals of the refinement made for the data collected at 15.5 GPa are shown in Fig. 2. The R-factors of the refinement are $R_p$ = 5.67% and $R_{wp}$ = 7.72%. The unit-cell parameters of scheelite-type $HoVO_4$ at 15.5 GPa are $a$ = 4.9916(16) Å and $c$ = 10.887(6) Å. Table I gives the atomic positions refined from this experiment for scheelite- and zircon-type $HoVO_4$.

In addition to the peaks assigned to zircon and scheelite $HoVO_4$ and Pt, there are a few weak peaks detected from 6.6 to 13.3 GPa. The most intense of them is located near 2θ = 12º and denoted by an asterisk in Fig. 1a. Consistent with the observed extra peaks is an orthorhombic structure (space group: *Pmmn*) with the same unit-cell parameters than $V_2O_5$ [38]. This happens at all the pressures where the extra peaks have been observed. Therefore, the possibility that a partial pressure-induced decomposition of $HoVO_4$ took place in run 1 appears as a probable hypothesis to explain our observations. Fortunately, the partial decomposition of the sample and the Bragg peaks associated to it do not preclude the identification of the zircon and scheelite structures at any pressure. The Bragg peaks of either one of these two structures are always the dominant peaks in all XRD patterns. Additionally, at 15.5 GPa the peaks assigned to $V_2O_5$ disappear. This fact is explained by the pronounced structural disorder induced by pressure in $V_2O_5$ [39], which leads to broadening and weakening of Bragg peaks in $V_2O_5$ and ultimately to its amorphization [39]. The cause of the partial decomposition of $HoVO_4$ is not clear, but it has been observed only in run 1. One possibility is that partial decomposition could be trigger by x-ray absorption which could induce photoelectric processes leading to the dissociation of $V_2O_5$ units from $HoVO_4$. Such a decomposition has been recently observed in ternary oxides in HP experiments [40], in particular when x-rays with wavelengths larger than 0.6 Å are used. After the study of Pravica *et al.* [40], we considered that in $HoVO_4$ the x-ray induced decomposition may be caused by the dissociation of $V_2O_5$ from the starting material. This dissociation has been suggested to be associated to a bond-related resonance of x-ray standing waves generated within the unit-cell [40]. According with this hypothesis, decomposition would be maximized for x-ray wavelengths close to half of the V-O bond length. This is the case of run 1 (0.6888 Å). Additionally, in run 1 the acquisition time was ~50 times longer than in run 2, which could have favored the observed decomposition. Further studies are needed to confirm this hypothesis and to check the x-ray wavelength influence on the induction of decomposition in HoVO$_4$ and related vanadates.

Upon further compression, beyond 15.5 GPa, we found that starting at 20.3 GPa the Bragg peaks of HoVO$_4$ considerably broadens. Additionally, extra weak peaks appear. In particular, in the strong (112) reflection of the scheelite phase develops a shoulder on the right-hand side becoming asymmetric. Similar changes have been observed in HP diffraction studies in other vanadates [12, 41]. These changes are consistent with a transformation from the scheelite structure to a monoclinic fergusonite structure (space group: *I2/a*, Z = 4) as previously observed at similar pressure in LuVO$_4$ and EuVO$_4$ among other vanadates [12, 42] and in related oxides [42 – 45]. The fergusonite structure of HoVO$_4$ remains stable up to 23.7 GPa, the highest pressure reached in run 1. The unit-cell parameters of fergusonite-type HoVO$_4$ at 23.7 GPa are *a* = 4.8331(34) Å, *b* = 10.765(6) Å, *c* = 5.043(4) Å, and β = 92.19(4)°. The residuals of the refinement made for the data collected at 23.7 GPa are shown in Fig. 2. The R-factors of the refinement are $R_p$ = 4.95% and $R_{wp}$ = 6.83%. Table I gives the atomic positions obtained for the fergusonite structure. Upon decompression the scheelite-fergusonite transition is reversible. This is shown in Fig. 1c, where it can be seen a series of XRD patterns measured during decompression. The pattern measured at 21.2 GPa can be assigned to the fergusonite structure. When reducing the pressure to 14.5 GPa the scheelite structure is recovered. Unfortunately, we do not have any measured XRD pattern in between these two, and thus the hysteresis of the transition cannot be accurately quantified. When totally releasing the force applied to the DAC a pressure of 0.8 GPa is achieved due to piston-cylinder friction in the DAC. At this pressure, the scheelite structure remains as a metastable phase as can be seen in Fig. 1c. The same phenomenon has been observed in related vanadates [12, 41]. The unit-cell parameters of metastable scheelite-type HoVO$_4$ at 0.8 GPa are *a* = 5.0440(12) Å and *c* = 11.159(5) Å. The residuals of the refinement made for scheelite-type HoVO$_4$ at 0.8 GPa are shown in Fig. 2. The R-factors of the refinement are $R_p$ = 5.33% and $R_{wp}$ = 7.61%. The obtained unit-cell parameters are comparable with those extracted by Stubican and Roy [21] from scheelite HoVO$_4$ at ambient pressure.

We would like to comment here that the scheelite-fergusonite transition has been previously observed in TbVO$_4$ at pressures ranging from 27 to 34 GPa [11, 15] depending upon the PTM used in the experiments. The use of different PTM could induce different deviatoric stresses within the pressure chamber of the DAC leading to conditions that range from quasi-hydrostatic to non-hydrostatic. This could strongly influence the HP structural sequence of materials even at low pressures [46, 47]. This influence is particularly notorious in scheelite-type oxides [48, 49]. Consequently, we decided to carry out a second experiment (run 2) in which a more hydrostatic PTM (Ar) was used. Fig. 3 shows a selection of XRD patterns measured in run 2. In these patterns all observed peaks can be assigned to HoVO$_4$, Ar, and Pt with the exception of two weak peaks that can be assigned to the gasket material. The more intense of these two peaks is identified in the figure with an asterisk. It is important to note that in this series of measurements the peaks of the sample (and also those of Pt and Ar) do not broaden considerably upon compression up to the highest pressure reached in the experiments. This fact indicates that experimental conditions do not deviate considerably from quasi-hydrostaticity [50]. In contrast, in run 1, beyond 8 GPa Bragg peaks gradually broaden, including those of Pt, which does not undergo any phase transition in the pressure range covered by our experiments [25]. In particular, the (111) peak of Pt (the most intense one) has a nearly constant full width at half maximum value of 0.15º up to 8 GPa; however this magnitude linearly increases with pressure reaching a value of 0.20º at 20 GPa. This fact indicates that a certain degree of non-hydrostaticity is present in run 1 beyond 8 GPa and therefore deviatoric stresses cannot be neglected in run 1 beyond this pressure.

In run 2, we found that the XRD patterns measured from ambient pressure up to 7.5 GPa can be assigned unequivocally to the zircon structure. This is illustrated in Fig. 3 by the patterns measured at 0.05 and 7.1 GPa. The Rietveld refinement and the residuals at 0.05 GPa are shown in the figure. The residuals of the refinement are R$_p$ = 2.95% and R$_{wp}$ = 4.42%. Similar residuals were obtained up to 7.5 GPa. The unit-cell parameters determined for zircon-type HoVO$_4$ at 0.05 GPa are $a$ = 7.123(5) Å and $c$ = 6.289(9) Å. The refined atomic positions are given in Table II. They agree with the parameters obtained from run 1 and those determined at ambient pressure [22]. In Fig. 3, it can be seen that in run 2 at 8.2 GPa new peaks emerge (denoted by $). These peaks are consistent with a scheelite-type structure. At 8.8 GPa, these peaks have gained in intensity while the zircon peaks have become weaker. The zircon peaks are undetectable at 9.2 GPa and only peaks assigned to scheelite-type HoVO$_4$, Pt, Ar, or gasket can be found at this pressure. Consequently, in the quasi-hydrostatic experiment, the onset of the transition occurs at a pressure ~4 GPa higher than in the non-hydrostatic experiment. On the other hand, the range of coexistence of both phases is reduced in the quasi-hydrostatic experiment. Rietveld refinements at pressures where pure scheelite or zircon plus scheelite are detected are also shown in Fig. 3. For the pattern measured at 18 GPa, the residuals of the refinement are R$_p$ = 2.99% and R$_{wp}$ = 4.65%. The unit-cell parameters determined for scheelite-type HoVO$_4$ at this pressure are $a$ = 4.890(5) Å and $c$ = 10.770(9) Å. Another difference to remark between run 1 and 2 is that in the quasi-hydrostatic experiment the scheelite-fergusonite transition is not observed up to 28 GPa while in run 1 it is obtained at 20.4 GPa. This fact is in agreement with observations made in TbVO$_4$ [11, 15] in which deviatoric stresses reduced the

transition pressure of the second transition by 7 GPa. On the other hand, as in run 1, in run 2 we also found the zircon-scheelite transition to be irreversible. At 0.1 GPa the unit-cell parameters of scheelite-type HoVO$_4$ are $a$ = 5.021(5) Å and c = 11.209(9) Å. They agree with values obtained in run 1 and reported in the literature [21]. The refined atomic positions are given in Table II.

*4.2. Ab initio calculations*

In order to theoretically study the possible HP phases of HoVO$_4$ we considered in addition to zircon several candidate structures. They included scheelite, fergusonite, monazite (a structure found under HP in zircon-type phosphates [51]), the orthorhombic structures (space group *Fddd* and *Imma*) found in HoVO$_4$ at low temperature [5, 6], and an orthorhombic structure with space group *Cmca* proposed as a post-scheelite structure in related vanadates [15]. Fig. 4 shows the enthalpy versus pressure for the different structures that have been considered. Our study indicates that the zircon-type is the structure of HoVO$_4$ with the lowest enthalpy at ambient pressure. The calculated structural parameters at ambient pressure and selected pressures for the different phases are given in Table III. They agree well with the experimental results.

Upon compression calculations predict the occurrence of a zircon-to-scheelite at 5.5 GPa. This first transition is in agreement with present and previous experiments [1]. It is a first-order transition that involves a large volume collapse ($\Delta V/V$ = 11 %). The tetragonal scheelite structure, as the zircon structure, consists of HoO$_8$ bisdisphenoids and VO$_4$ tetrahedra [52]. The structural relation between both structures has been nicely described by Nyman *et al.* [52]. The transition is first-order reconstructive and large kinetic barrier are associated to it [8]. This could justify that calculations find a transition pressure slightly smaller than quasi-hydrostatic experiments. In the calculations we also found that most of the structures considered as potential HP phases are not energetically competitive with zircon and scheelite. In particular, the orthorhombic structures with space groups *Fddd* and *Imma* (subgroups of *I4$_1$/amd*), which can be considered distortions of zircon, at all considered pressures after optimization reduce to the zircon structure. The same happens with fergusonite (a monoclinic distortion of scheelite) which at all pressures reduces to the scheelite structure. In addition, we found that the monazite structure is not thermodynamically stable at any pressure. This is consistent with the fact that it has been observed only in vanadates with rare earths with a large ionic radius, like La, Ce, and Nd [9 – 11]. On the other hand, when pressure is 32 GPa or larger, we have found a structure that becomes thermodynamically more stable than zircon and scheelite. This structure has an orthorhombic symmetry, with space group *Cmca* and double number of atoms in the unit cell (Z = 8) than zircon and scheelite. The structural details of this structure are given in Table III. In the orthorhombic structure, the Ho and V cations are surrounded by eleven and seven oxygen atoms, respectively; i.e. the coordination number increases for both cations. The transition to the post-scheelite structure is predicted to occur at a pressure larger than the maximum pressure covered in present experiments. The existence of a similar structure as post-scheelite phase in TbVO$_4$ [15] suggests that the present predictions deserve to be tested by future experiments.

We would like to note here that since only a limited number of structures were analyzed as candidate HP phases in HoVO$_4$, we cannot exclude the existence of other HP structures not considered in our calculations. However, the presence of an orthorhombic structure belonging to space group *Cmca* at extreme pressure is fully consistent with crystal chemistry arguments. It has been shown that an analogue can be made between the zircon and rutile structure (TiO$_2$) and between their HP phases [53]. In particular, the rutile structure consists of infinite rectilinear rods of edge-sharing TiO$_6$ octahedra parallel to the *c*-axis, linked by corner sharing to the octahedra in identical corner rods. If Ti is alternatively substituted by Ho and V atoms and they are shifted in each rod then the zircon structure is obtained. Therefore zircon can be considered a superstructure of rutile. Under compression rutile transforms to the $\alpha$-PbO$_2$-type structure [54] and scheelite can be thought as distorted superstructure of $\alpha$-PbO$_2$. In particular, scheelite is related to the $\alpha$-PbO$_2$ structure in a way that is analogous to the relationship between zircon and rutile. Additionally, the orthorhombic post-scheeliite structure predicted for HoVO$_4$ is related to the pyrite structure, the last link of the HP sequence of rutile-type oxides [54]. We think these arguments provide additional support to the zircon-scheelite-orthorhombic structural sequence predicted by our calculations under hydrostatic conditions.

In agreement with the quasi-hydrostatic experiments of run 2, our calculations did not find that scheelite transforms into ferguson. In related compounds, it has been reported that deviatoric stresses could affect the HP structural sequence [11, 48]. To check if this is the case of HoVO$_4$, we carried out calculations under non-hydrostatic conditions. The result of the calculations is that deviatoric stresses of 1.2 GPa will be enough to induce the transition from scheelite to ferguson when the hydrostatic component of the stress tensor is larger than 20 GPa. This result is in agreement with the findings of run 1, confirming that to trigger the scheelite-ferguson, the existence of deviatoric stresses is fundamental. For the sake of comparison the *ab initio* results for the ferguson structure around 21 GPa obtained imposing non-hydrostatic conditions are given in Table III. It resembles very much the structure found in the experiments.

*4.3. Equations of state*

From the experiments we extracted the pressure dependence of the unit-cell parameters of the different phases of HoVO$_4$ as shown in Fig. 5. Fig. 6 outlines the pressure dependence of the unit-cell volume. In Fig. 5, one can see that the lattice parameters of the zircon structure are less compressible in run 1 than in run 2. The difference is more noticeable in the *c* axis. In addition, the quasi-hydrostatic run 2 agrees better with calculations than run 1. Differences are larger for the scheelite structure, in particular beyond 10 GPa. At these pressures, run 1 deviates from the evolution followed by the unit-cell parameters in run 2 and from the theoretical results. Clearly, the zircon and scheelite phases of $HoVO_4$ are less compressible under non-hydrostatic compression than under quasi-hydrostatic compression. For completeness, in Fig. 5 we have included the results for fergusonite phase as well. The increase of the β angle under compression (see inset of Fig. 5) and the increase of the difference between *a* and *c* axes indicate that the monoclinic distortion of fergusonite gradually increases under pressure. This phenomenon should be reflected in an enhancement of the spontaneous strains that characterize the distortion caused by the transformation from scheelite to fergusonite [55].

The obtained P-V data, shown in Fig. 6, are fitted using a third-order Birch-Murnaghan (BM) EOS to obtain the ambient pressure bulk modulus $B_0$ and its pressure derivative $B_0'$ as well as the unit-cell volume $V_0$ [56]. The EOS parameters are given in Table IV together with the calculated values. For calculations a fourth-order BM EOS [56] was used. Therefore, for comparison we provide for experiments the implied value of the second pressure derivative of the bulk modulus $B_0''$ [57]. To fit the experimental P-V results we used for run 2 all the pressure range covered by the experiments. For run 1 we only used results for P ≤ 10 GPa to minimize the influence of deviatoric stresses in the fits. We assumed the three EOS parameters as fitting parameters. The only exception is scheelite $HoVO_4$ from run 1. In this case since the pressure range is constrained we have only ten data points. Thus to reduce the number of fitting parameters we assume for $V_0$ a fixed value, the one obtained from run 2. The reported values for the EOS parameters of the different phases of $HoVO_4$ are similar to those previously found in isomorphic orthovanadates [8 – 12, 41]. Additionally, in both experiments and in calculations, the scheelite phase has a bulk modulus 12% larger than the zircon phase. This is in agreement with the large volume collapse associated to the zircon-scheelite transition and to the associated increase of packing efficiency of the scheelite structure. In Table IV it can be seen that the quasi-hydrostatic experiments agree better with theory than the experiments done using ME as PTM. In particular, the calculations gave a bulk modulus 10% smaller that the quasi-hydrostatic experiment (for both the zircon and scheelite structures). This is typical of DFT calculations [58] and is consistent with the fact that calculations slightly overestimate the unit-cell volume; in particular for zircon- and scheelite-type oxides. On the other hand, the ME experiment gave a bulk modulus 12% larger than the Ar experiment. This suggests that deviatoric stresses causes a reduction of bulk compressibility (increase of bulk modulus) as has been already found in other compounds [48], where as large as 30% differences have been observed in the bulk modulus when different PTM were used in the experiments.

Once discussed the EOS of zircon and scheelite, we would like to comment on the compressibility of the orthorhombic HP phase. According to calculations, the transition from scheelite to this structure implies a volume collapse of approximately 8%, and a cation coordination increase as described above. However, in spite of the density increase, the bulk modulus of the orthorhombic phase is slightly smaller than in the scheelite phase. The same fact has been predicted for the same scheelite-orthorhombic transition in TbVO$_4$ [15]. It has been explained as manifestation of localized-to-delocalized electronic transition [59]. In particular, the known *f*-electron delocalization induced by pressure in lanthanides [60 – 62] might weaken some of the Ho–O bonds under pressure in the orthorhombic HP phase leading to the predicted decrease of the bulk modulus after the second phase transition. To conclude this section, we would like to add that none of the structures of HoVO$_4$ shows anomalous positive values for the second pressure derivative of the bulk modulus (see Table IV). Therefore, the rate at which the four phases become stiffer decreases with increasing pressure.

*4.4. Interatomic distances*

As commented above, the zircon-scheelite transition is a first-order transition which involves important changes in bond distances, however, the cation coordination is not modified at the transition. It has been shown before that the macroscopic behavior of vanadates under compression can be better understood studying the effects of pressure on interatomic bond distances. With this aim, from our calculations we obtained the pressure dependence for the cation–oxygen distances for the zircon and scheelite phases. Results are displayed in Fig. 7. For both phases, the interatomic distances vary smoothly with pressure. In the zircon phase, the Ho–O distances display a stronger change with pressure than the V–O distances. In addition, the distortion of the HoO$_8$ dodecahedra is enhanced under compression because the short Ho-O distance is more compressible than the long Ho-O distance. In going from zircon to scheelite at the calculated transition pressure, the V–O distance increases by about 1%, whereas the two Ho–O distances becomes more similar in the scheelite phase than in the zircon phase. The evolution of Ho-O distances in scheelite HoVO$_4$ implies a regularization of the HoO$_8$ dodecahedra with the increase of pressure. The variation of interatomic distances in both structures compares qualitatively well with the earlier reported pressure variation for related vanadates [8 – 11]. Note that Ho-O bonds are more compressible than V-O bonds, which indicates that the large polyhedral units associated with the Ho atom are responsible for most of the volume change induced by pressure, a typical behavior of zircons and scheelites [44, 51]. To conclude, we will only add that in spite of the density increase associated to the scheelite-orthorhombic structure transition, in the orthorhombic phase the bond distances are enlarged as a consequence of the increase of the coordination number. In particular, at 44 GPa the bond distances in the orthorhombic structure are: V-O = 1.7562 Å, 1.8081 Å, 1.8838 Å (x2), 1.9540 Å (x2), and 2.2535 Å and Ho-O = 2.1543 Å (x2), 2.2257 Å, 2.2606 Å (x2), 2.3536 Å (x2), 2.4302 Å (x2), and 2.6432 Å (x2).

## 5. Conclusions

We performed room-temperature angle-dispersive XRD measurements on $HoVO_4$ up to 28 GPa under both quasi-hydrostatic and non-hydrostatic conditions. In both cases the irreversible zircon-scheelite transition was found at 8.2 and 4.5 GPa, respectively. In the second case, a second (reversible) transition is detected at 20.4 GPa. This transition is from scheelite to fergusonite. In addition, a partial decomposition of $HoVO_4$ was observed. We believe, it could be triggered by x-ray absorption when a large x-ray wavelength is used. The structure of the different phases has been refined and their equations of state (EOS) determined. *Ab initio* calculations have been also carried out. They fully agree with the quasi-hydrostatic experiments and show that deviatoric stresses are needed to stabilize the fergusonite structure beyond 20 GPa. Calculations also suggest the possibility of another phase transition at 32 GPa, a pressure higher than the maximum pressure reached in our experiments. The consistency between the values yielded for transition pressures and EOS parameters by theory and experiments is quite good.

**Acknowledgements**

This work was supported Spanish MINECO under projects MAT2010-21270-C04-01/03 and MALTA Consolider Ingenio CSD2007-00045. Supercomputer time has been provided by the Red Española de Supercomputación (RES) and the MALTA cluster. We thank ALBA and Elettra synchrotrons for providing beam-time for the XRD experiments. Alka B. Garg acknowledges the DST of India for travel support and Italian Government for hospitality at Elettra. She also acknowledges fruitful discussions with Dr. Surinder M. Sharma.

**References**

[1] Y.-Z. Cheng, S. Li, L. Li, Z.W. Men, Z L. Li, C.-L.Sun, and M. Zhou, *Acta Phys. Sin.* **62**, 246101 (2013).

[2] Z. Huang, L. Zhang, and W. Pan, *Inorg. Chem.* **51**, 11235 (2012).

[3] S.P. Shafi, M.W. Kotyk, L.M.D. Cranswick, V.K. Michaelis, S. Kroeker, and M. Bieringer, *Inorg. Chem.* **48**, 10553 (2009).

[4] Y. Zhang, G. Li, X. Yang, H. Yang, Z. Lu, and R. Chen, *J. Alloys Comp.* **551**, 544 (2013).

[5] G. Will and W. Schäfer, *J. Phys. C: Solid State Phys.* **4**, 811 (1971).

[6] K. Kirschbaum, A. Martin, D.A. Parish, and A.A. Pinkerton, *J. Phys. Condens. Matter* **11**, 4483 (1999).

[7] D. Errandonea, O. Gomis, B. Garcia-Domene, J. Pellicer-Porres, V. Katari, S.N. Achary, A.K. Tyagi, and C. Popescu, *Inorg. Chem.* **52**, 12709 (2013).

[8] W. Paszkowicz, O. Ermakova, J. López-Solano, A. Mujica, A. Muñoz, R. Minikayev, C. Lathe, S. Gierlotka, I. Nikolaenko, and H. Dabkowska, *J. Phys. Condens. Matter.* **26**, 025401 (2014).

[9] D. Errandonea, C. Popescu, S.N. Achary, A.K. Tyagi, and M. Bettinelli, *Materials Research Bulletin* **50**, 279 (2014).

[10] A.B. Garg, K.V. Shanavas, B.N. Wani, and S.M. Sharma, *J. Sol. State Chem.* **203**, 273 (2013).

[11] D. Errandonea, R.S. Kumar, S.N. Achary, and A.K. Tyagi, *Phys. Rev. B* **84**, 214121 (2011).

[12] D. Errandonea, R. Lacomba-Perales, J. Ruiz-Fuertes, A. Segura, S.N. Achary, and A.K. Tyagi, *Phys. Rev. B* **79**, 184104 (2009).

[13] V. Panchal, D. Errandonea, A. Segura, P. Rodriguez-Hernandez, A. Muñoz, S. Lopez-Moreno, and M. Bettinelli, *J. Appl. Phys.* **110**, 043723 (2011).

[14] S.J. Duclos, A. Jayaraman, G.P. Espinosa, A.S. Cooper, and R.G. Maines, *J. Phys. Chem. Solids* **50**, 769 (1989).

[15] D. Errandonea, F.J. Manjon, A. Muñoz, P. Rodriguez-Hernandez, V. Panchal, S.N. Achary, and A.K. Tyagi, *J. Alloys Comp.* **577**, 327 (2013).

[16] D. Errandonea, S.N. Achary, J. Pellicer-Porres, and A.K.Tyagi, *Inorg. Chem.* **52**, 5464 (2013).

[17] N.N. Patel, A.B. Garg, S.Meenakshi, B.N.Wani, and S.M. Sharma, *AIP Conference Proceedings.* **1349**, 99 (2011).

[18] R. Rao, A.B.Garg, T.Sakuntala, S.N. Achary, and A.K.Tyagi, *J. Solid State Chem.* **182**, 1879 (2009).

[19] Z. Huang, L. Zhang, and W. Pan, *J. Solid State Chem.* **205**, 97 (2013).

[20] S. Lopez-Moreno and D.Errandonea, *Phys. Rev. B* **86**, 104112 (2012).

[21] V.S. Stubican and R. Roy, *Zeitsch. Fúr Kristallog.* **119**, 90 (1963).

[22] B.C. Chakoumakos, M.M. Abraham, and L.A. Boatner, *J. Solid State Chem.* **109**, 197 (1994).

[23] D. Errandonea, R. Boehler, S. Japel, M. Mezouar, and L. R. Benedetti, *Phys. Rev. B* **73**, 092106 (2006).

[24] S. Klotz, J. C. Chervin, P. Munsch, and G. L. Marchand, *J. Phys. D: Appl. Phys.* **42**, 075413 (2009).

[25] A.Dewaele, P.Loubeyre, and M.Mezouar, *Phys. Rev. B* **70**, 094112 (2004).

[26] F. Fauth, I. Peral, C. Popescu, and M. Knapp, *Powder Diffraction* **28**, S360 (2013).

[27] A. P. Hammersley, S. O. Svensson, M. Hanfland, A. N. Fitch, and D. Häusermann, *High Pressure Research* **14**, 235 (1996).

[28] W. Kraus and G. Nolze, *J. Appl. Crystallogr.* **29**, 301 (1996).

[29] A. C. Larson and R. B. von Dreele, *LANL Report* 86-748 (2004).

[30] P. Hohenberg, W. Kohn, *Phys. Rev.* **136**, B864 (1964)

[31] G. Kresse and J. Furthmüller, *Phys. Rev. B* **54**, 11169 (1996).

[32] G. Kresse and D. Joubert, *Phys. Rev. B* **59**, 1758 (1999).

[33] P. E. Blochl, *Phys. Rev. B* **50**, 17953 (1994).

[34] J. Hafner, *J. Comput. Chem.* **29**, 2044 (2008).

[35] C. J. Pickard, B.Winkler, R. K. Chen, M. C. Payne, M. H. Lee, J. S. Lin, J. A. White, V.Milman, and D. Vanderbilt, *Phys. Rev. Lett.* **85**, 5122 (2000).

[36] J. P. Perdew, K. Burke, and M. Ernzerhof, *Phys. Rev. Lett.* **77**, 3865 (1996).

[37] A. Mujica, A. Rubio, A. Muñoz, and R. J. Needs, *Rev. Mod. Phys.* **75**, 863 (2003).

[38] I. Loa, A. Grzechnik, U. Scharz, K. Syassen, M. Hanfland, and R.K. Kremer, *J. Alloys Comp.* **317-318**, 103 (2001).

[39] A.K. Arora, T. Sato, T. Okada, and T. Yagi, *Phys. Rev. B* **85**, 094113 (2012).

[40] M.Pravica, L. Bai, D. Sneed, and C. Park, *J. Phys. Chem. A* **117**, 2302 (2013).

[41] R Mittal, A.B. Garg, V.Vijayakumar, S.N. Achary, A.K.Tyagi, B.K. Godwal, E.Busetto, A. Lausi, and S.L. Chaplot, *J. Phys.: Condens. Matter* **20**, 075223 (2008).

[42] D. Errandonea, J. Pellicer-Porres, F. J. Manjón, A. Segura, Ch. Ferrer-Roca, R. S. Kumar, O. Tschauner, P. Rodríguez-Hernández, J. López-Solano, S. Radescu, A. Mujica, A. Muñoz, and G. Aquilanti, *Phys. Rev. B* **72**, 174106 (2005).

[43] D. Errandonea, *Phys. Stat. Sol. B* **242**, R125 (2005).

[44] D. Errandonea, R.S. Kumar, X. Ma, and C. Tu, *Journal of Solid State Chemistry* **181**, 355 (2008).

[45] D. Errandonea, D. Santamaria-Perez, S.N. Achary, A.K. Tyagi, P. Gall, and P. Gougeon, *J. Appl. Phys.* **109**, 043510 (2011).

[46] Y. Meng, D. J. Weidner, and Y. Fei, *Geophys. Res. Lett.* **20**, 1147 (1993).

[47] D. Errandonea, Y. Meng, M. Somayazulu, D. Häusermann, *Physica B* **355**, 116 (2005).

[48] O. Gomis, J. A. Sans, R. Lacomba-Perales, D. Errandonea, Y. Meng, J. C. Chervin, and A. Polian, *Phy. Rev. B* **86**, 054121 (2012).

[49] D. Errandonea, L. Gracia, R. Lacomba-Perales, A. Polian, and J. C. Chervin, *J. Appl. Phys.* **113**, 123510 (2013).

[50] K. Takemura, *J. Appl. Phys.* **89**, 662 (2001).

[51] R. Lacomba-Perales, D. Errandonea, Y. Meng, M. Bettinelli, *Phys. Rev. B* **81**, 064113 (2010).

[52] H. Nyman, B.G. Hyde, and S. Andersson, *Acta Cryst. B* **40**, 441 (1984).

[53] D. Errandonea and F.J. Manjon, *Progress in Materials Science* **53**, 711 (2008).

[54] V.P. Prakapenka, G. Shen, L.S. Dubrovinsky, M.L. Rivers, and S.R. Sutton, *J. Phys. Chem. Solids* **65**, 1537 (2004).

[55] D. Errandonea, *EPL* **77**, 56001 (2007).

[56] F. Birch, *J. Geophys. Res.* **57**, 227 (1952).

[57] D. Errandonea, Ch. Ferrer-Roca, D. Martínez-Garcia, A. Segura, O. Gomis, A. Muñoz, P. Rodríguez-Hernández, J. López-Solano, S. Alconchel, and F. Sapiña, *Phys. Rev. B* **82**, 174105 (2010).

[58] D. Errandonea, R. Kumar, J. Lopez-Solano, P. Rodrıguez-Hernandez, A. Muñoz, M. G. Rabie, and R. Saez Puche, *Phys. Rev. B* **83**, 134109 (2011).

[59] F. Rivadulla, M. Bañobre-Lopez, C.X. Quintela, A. Piñeiro, V. Pardo, D. Baldomir, M.A. Lopez-Quintela, J. Rivas, C.A. Ramos, H. Salva, J.S. Zhou, J.B. Goodenough, *Nat. Mater.* **8**, 947 (2009).

[60] W.A. Grosshans, W.B. Holzapfel, *Phys. Rev. B* **45**, 6603 (1992).

[61] D. Errandonea, R. Boehler, M. Ross, *Phys. Rev. Lett*. **85**, 3444 (2000).

[62] Y.Y. Boguslavskii, V.A. Goncharova, G.G. Il'ina, *Zh. Eksp, Teor. Fiz.* **91**, 1735 (1986).

**Table I.** Refined atomic positions from XRD patterns measured in run 1 for zircon (top), scheelite (center), and fergusonite (bottom) HoVO$_4$ at 0.6, 15.5, and 23.7 GPa, respectively.

| Atom | Site | x | y | z |
|---|---|---|---|---|
| Ho | 4a | 0 | 3/4 | 1/8 |
| V | 4b | 0 | 1/4 | 3/8 |
| O | 16h | 0 | 0.4608(32) | 0.193(4) |
| Ho | 4b | 0 | 1/4 | 5/8 |
| V | 4a | 0 | 1/4 | 1/8 |
| O | 16f | 0.326(12) | 0.265(16) | 0.0948(27) |
| Ho | 4e | 1/4 | 0.626(3) | 0 |
| V | 4e | 1/4 | 0.129(7) | 0 |
| O | 8f | 0.996(18) | 0.210(11) | 0.895(20) |
| O | 8f | 0.067(11) | 0.438(6) | 0.746(40) |

**Table II.** Refined atomic position from XRD patterns measured in run 2 for zircon (top) and scheelite (bottom) at 0.05 and 0.1 GPa, respectively.

| Atom | Site | x | y | z |
|---|---|---|---|---|
| Ho | 4a | 0 | 3/4 | 1/8 |
| V | 4b | 0 | 1/4 | 3/8 |
| O | 16h | 0 | 0.4345(9) | 0.2005(8) |
| Ho | 4b | 0 | 1/4 | 5/8 |
| V | 4a | 0 | 1/4 | 1/8 |
| O | 16f | 0.2638(5) | 0.1975(8) | 0.0860(8) |

**Table III.** Calculated structural parameters for zircon, scheelite, and orthorhombic HoVO$_4$ at ambient pressure, 14.9 GPa, and 44 GPa, respectively and for fergusonite HoVO$_4$ obtained under non-hydrostatic conditions at 21 GPa.

| Atom | site | x | y | z |
|---|---|---|---|---|
| \multicolumn{5}{c}{Zircon P = 0 GPa, $a$ = 7.169 Å, $c$ = 6.304 Å} | | | | |
| Ho | 4a | 0 | 0.75 | 0.125 |
| V | 4b | 0 | 0.25 | 0.375 |
| O | 16h | 0 | 0.43514 | 0.19959 |
| \multicolumn{5}{c}{Scheelite P = 14.89 GPa, $a$ = 5.055 Å, $c$ = 11.248 Å} | | | | |
| Ho | 4b | 0 | 0.25 | 0.625 |
| V | 4a | 0 | 0.25 | 0.125 |
| O | 16f | 0.255866 | 0.10358 | 0.04442 |
| \multicolumn{5}{c}{Orthorhombic (*Cmca*) P = 44 GPa, $a$ = 7.159 Å, $b$ = 11.902 Å, $c$ = 4.857 Å} | | | | |
| Ho | 8e | 0.25 | 0.15979 | 0.25 |
| V | 8f | 0 | 0.41179 | 0.22449 |
| O | 8f | 0 | 0.08458 | 0.08590 |
| O | 8d | 0.34340 | 0 | 0 |
| O | 8f | 0 | 0.21097 | 0.50505 |
| O | 8e | 0.25 | 0.34678 | 0.25 |
| \multicolumn{5}{c}{Fergusonite P =21 GPa (non-hydrostatic), $a$ = 4.885 Å, $b$ = 10.712 Å, $c$ = 5.016 Å, $\beta$ = 91.54°} | | | | |
| Ho | 4e | 0.25 | 0.62558 | 0.00 |
| V | 4e | 0.25 | 0.11953 | 0.0 |
| O | 8f | 0.99786 | 0.20942 | 0.84743 |
| O | 8f | 0.08651 | 0.46060 | 0.75346 |

**Table IV.** EOS parameters for different structures determined from experiments and calculations. For the experimental values B0'' is the implied value and the used PTM is indicated.

| Phase | PTM | $V_0$ (Å$^3$) | $B_0$ (GPa) | $B_0'$ | $B_0''$ (GPa$^{-1}$) |
|---|---|---|---|---|---|
| zircon | ME | 320.5(9) | 160(15) | 5(2) | -0.037 |
| zircon | Ar | 319.1(1) | 143(3) | 4.7(9) | -0.036 |
| zircon | Theory | 324.6 | 128.7 | 5.4 | -0.113 |
| scheelite | ME | 282.5 | 180(15) | 5.5(9) | -0.042 |
| scheelite | Ar | 282.5(2) | 160(3) | 4.3(9) | -0.027 |
| scheelite | Theory | 287.8 | 144.1 | 4.6 | -0.074 |
| Cmca | Theory | 514.79 | 133.87 | 4.17 | -0.066 |

**Figure captions**

**Figure 1.** Selection of XRD patterns measured during compression in run 1 (ME used as PTM). Pressures are indicated. The Bragg peaks of Pt are identified and the indices of most relevant peaks of zircon and scheelite phases are shown. The symbol $ identify the most intense peak of scheelite in the pressure range where this phase coexists with zircon. The symbol * identifies the most intense peak associated to $V_2O_5$.

**Figure 2.** Rietveld refinements for different phases of $HoVO_4$; zircon (0.6 GPa), scheelite (15.5 GPa), fergusonite (23.7 GPa), a recovered scheelite (0.8 GPa). Dots correspond to experiments, the refinements (residuals) are shown with red (black) solid lines. Bragg peaks positions of $HoVO_4$ (Pt) are indicated by black (blue) ticks.

**Figure 3.** (top) Selection of XRD patterns measured in run 2 (Ar used as PTM). Pressures are indicated. Ar and Pt peaks are identified. The symbol $ identifies most intense peaks of scheelite when this is the minority phase. The symbol * identifies the most intense gasket peak. (bottom) Rietveld refinements at selected pressures. Experiments are shown as dots. Refinements (residuals) are shown as red (black) solid lines. Ticks indicate the position of Bragg peaks of zircon and scheelite. At 0.05 and 18 GPa they are labeled.

**Figure 4.** Enthalpy difference as a function of pressure curves showing the phase transitions here reported. The zircon phase has been taken as a reference.

**Figure 5.** Unit-cell parameters versus pressure. Solid symbols: compression. Empty symbols: decompression. We show results from run 1 (ME) and run 2 (Ar) as well as *ab initio* results (solid lines). Different symbols are identified within the figure. The inset shows the evolution of the β angle in the fergusonite phase (run 1).

**Figure 6.** Unit-cell volume versus pressure. We show results from run 1 (ME) and run 2 (Ar) as well as *ab initio* results (solid lines) and the EOS fitted from run 2 (dashed line). Different symbols are identified within the figure.

**Figure 7.** Pressure dependence of Ho-O and V-O distances in the zircon and scheelite structures of $HoVO_4$.

**Figure 1**

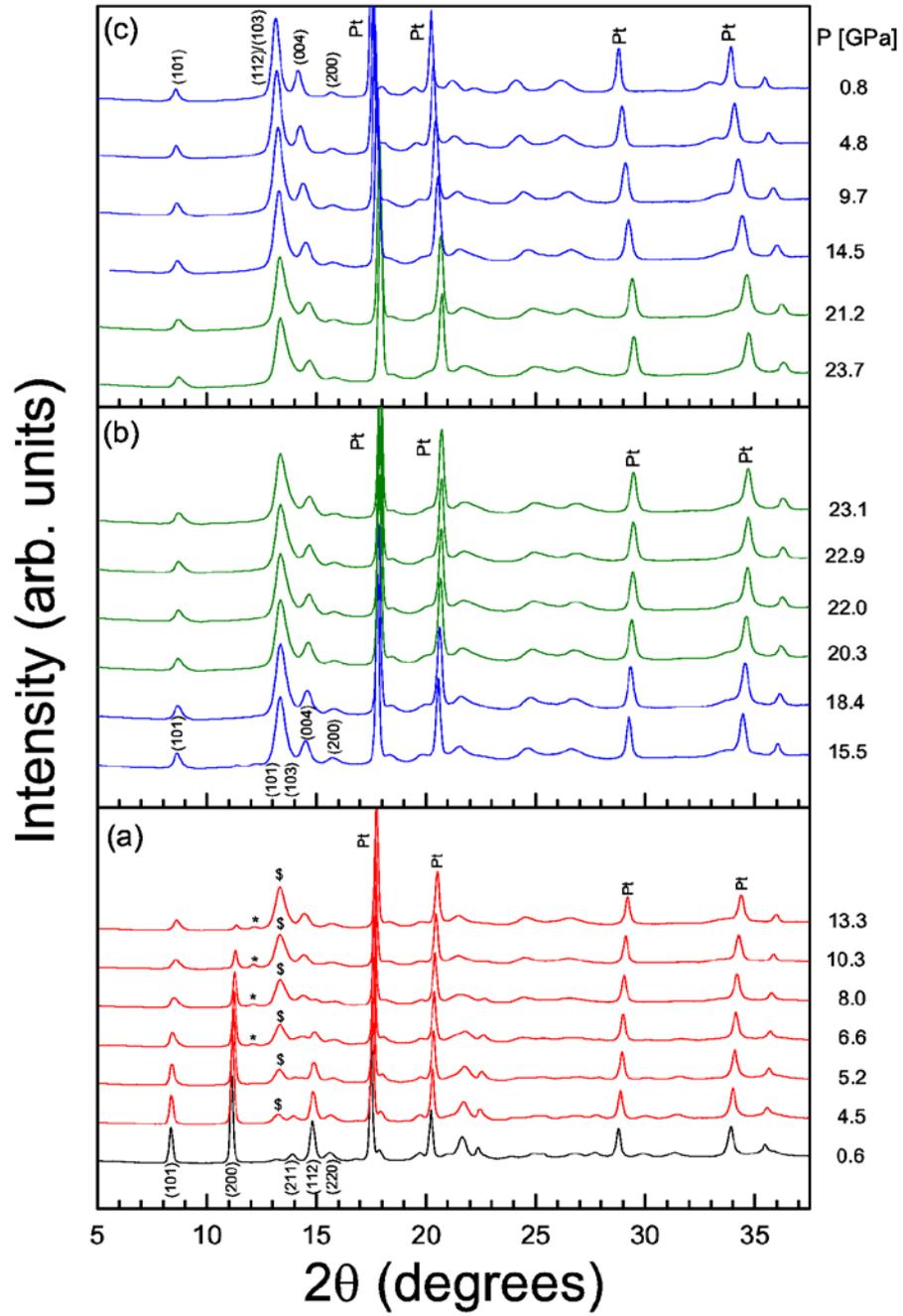

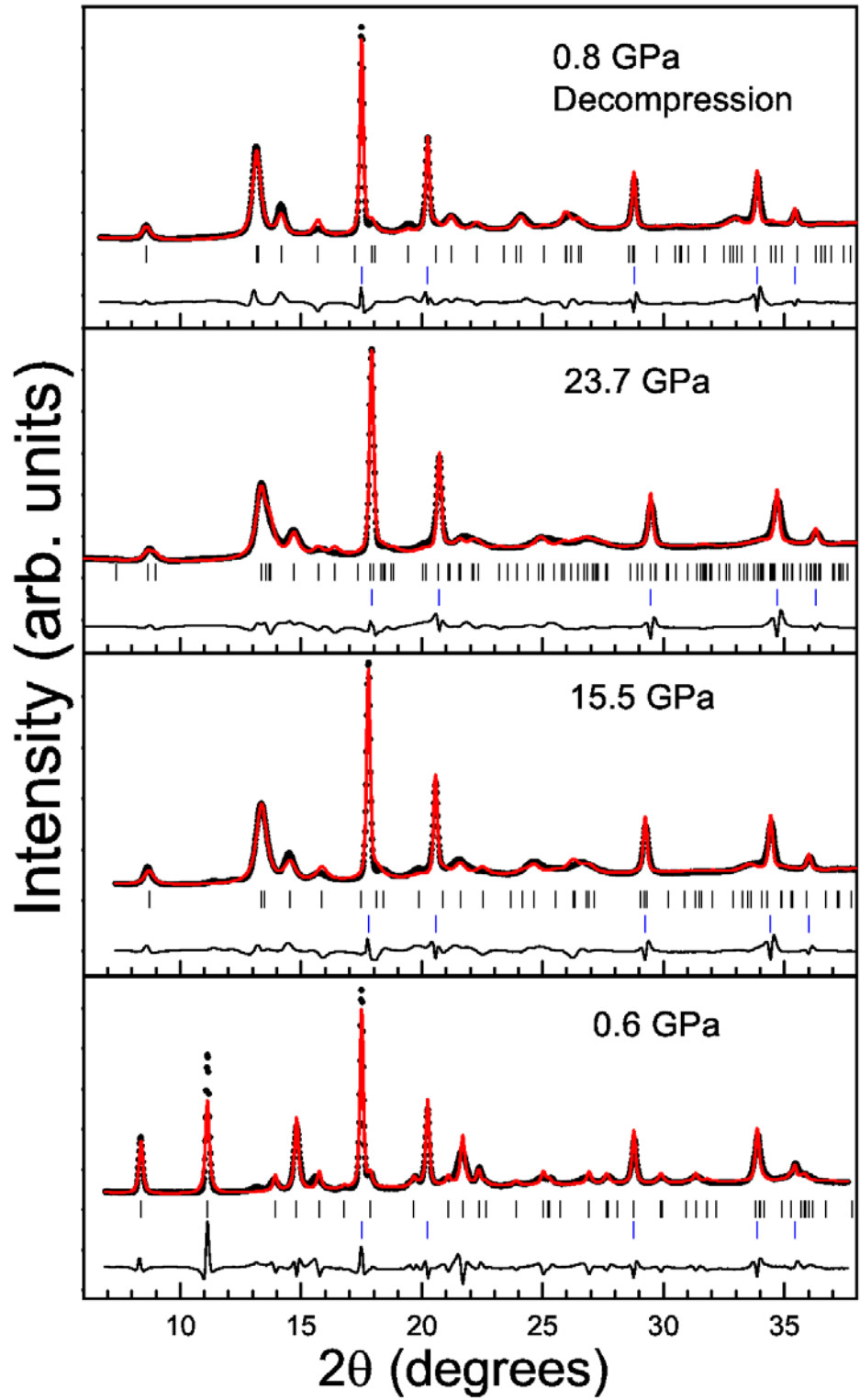

**Figure 3**

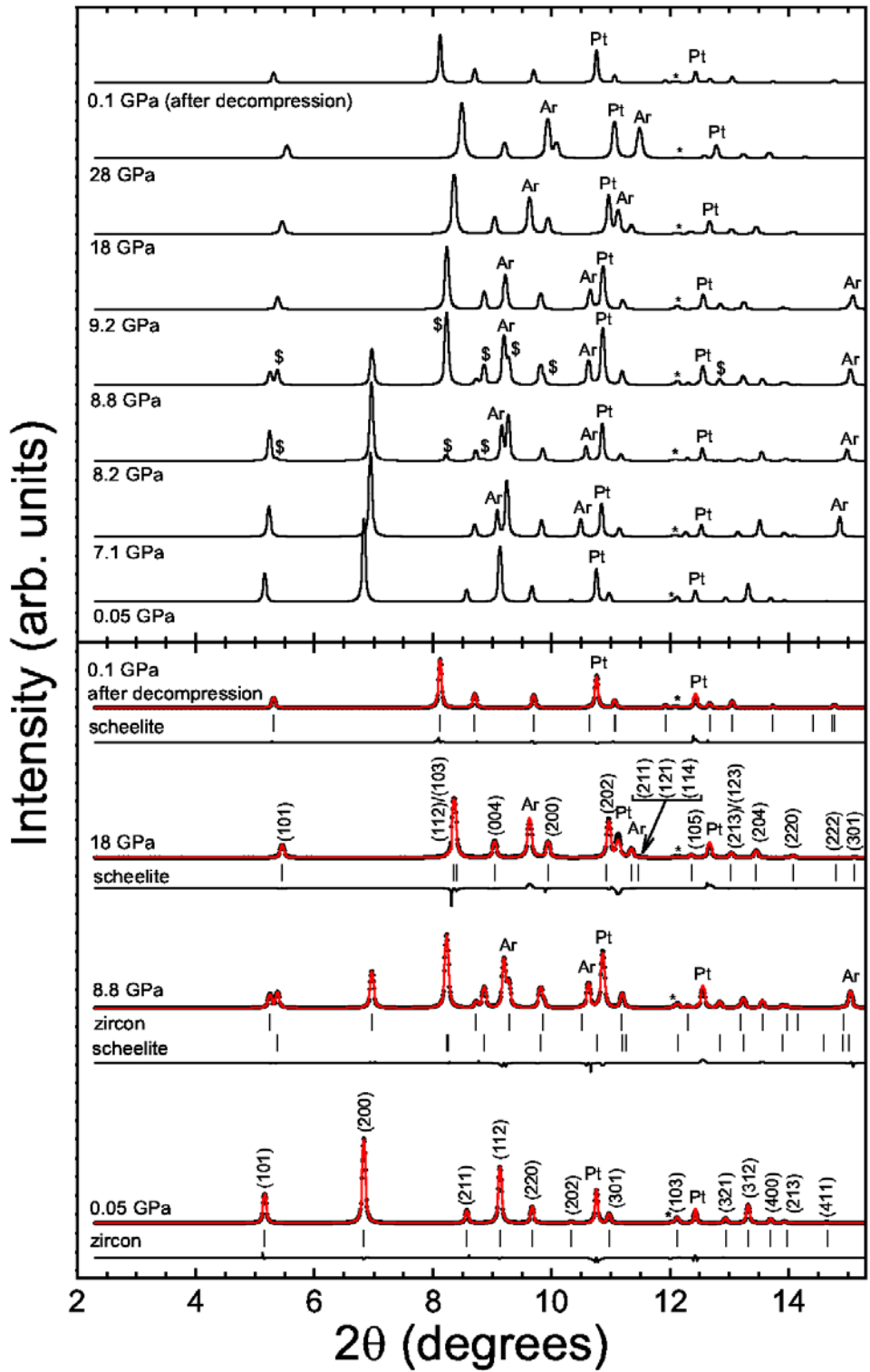

**Figure 4**

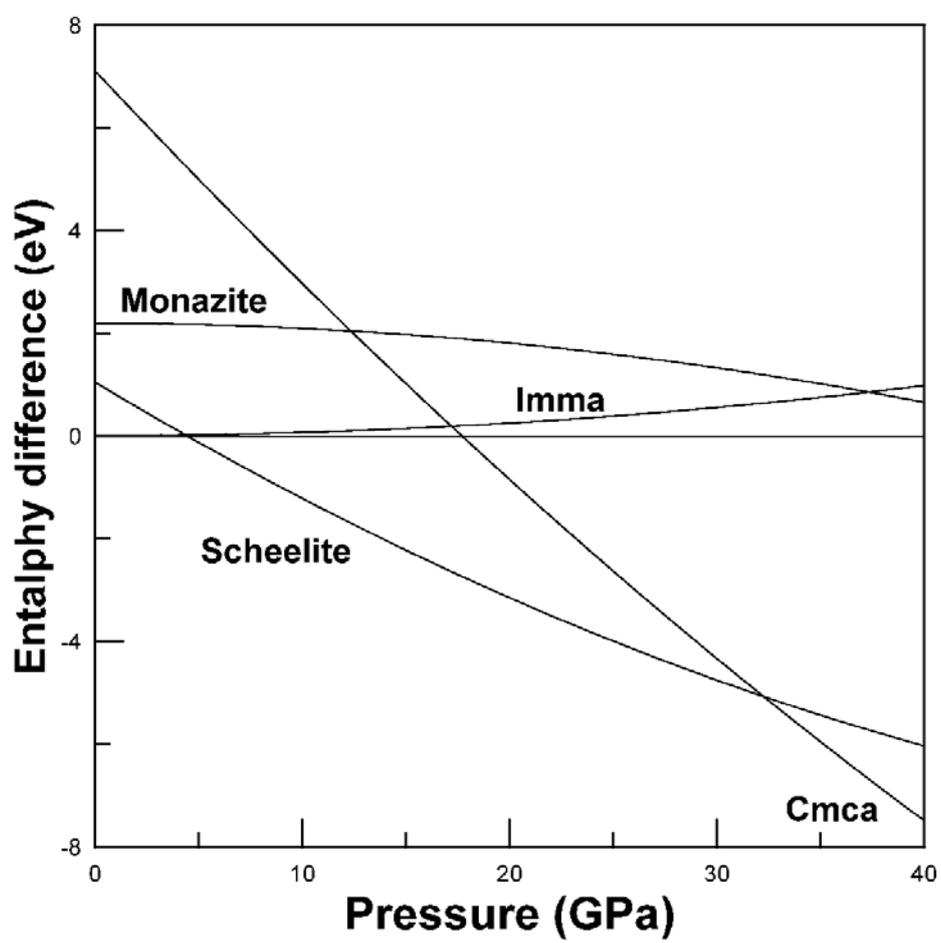

**Figure 5**

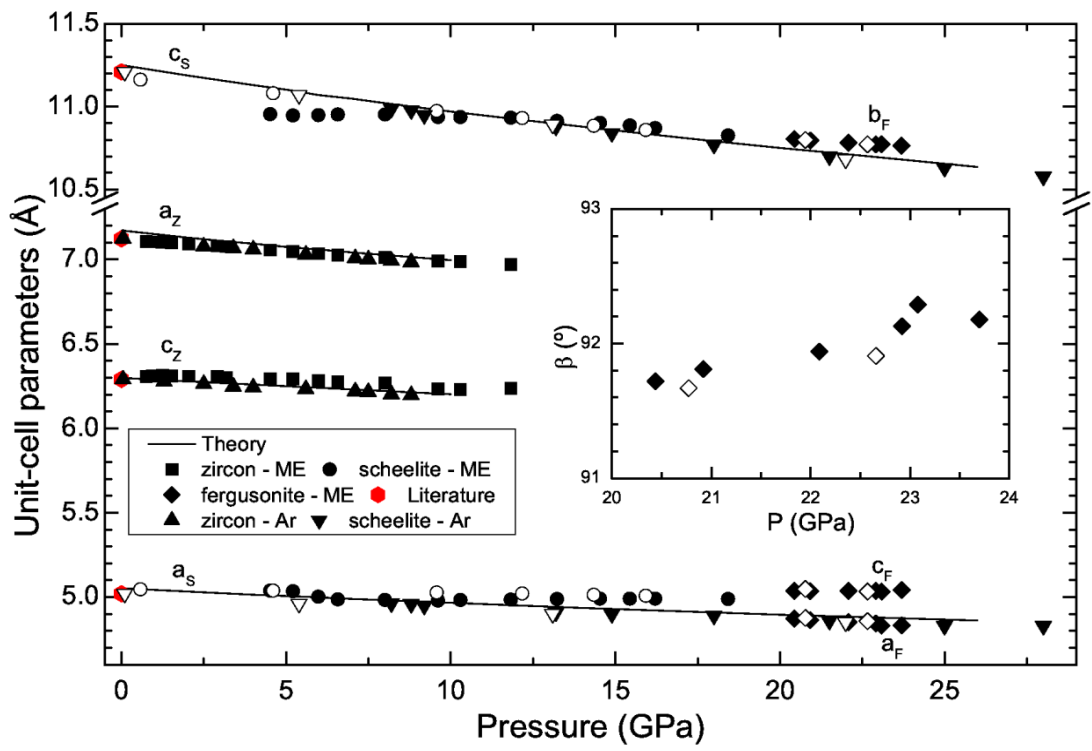

**Figure 6**

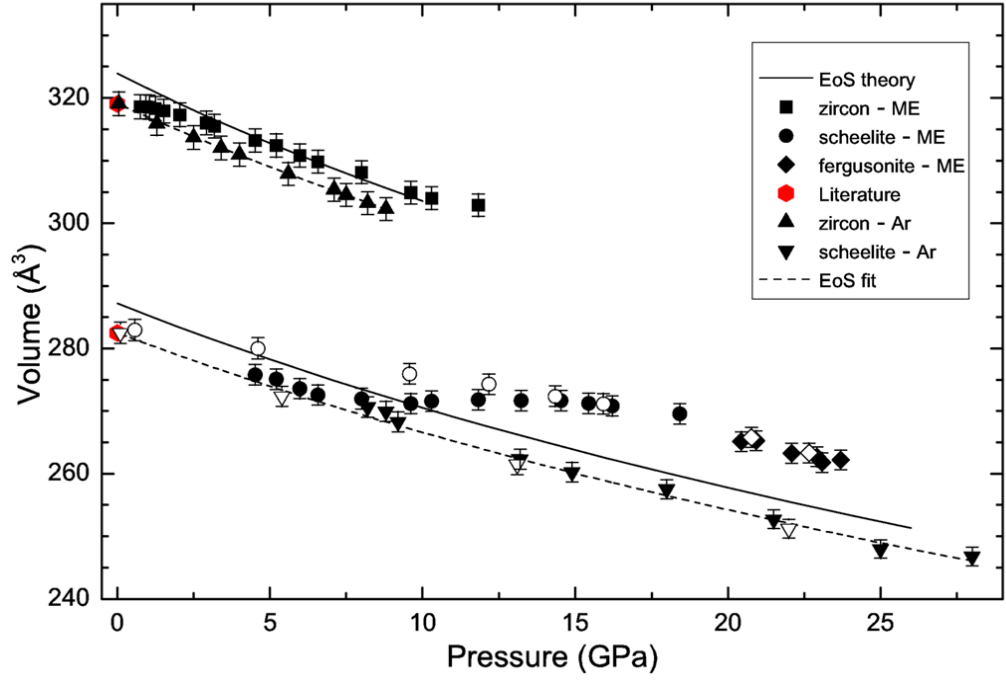

**Figure 7**

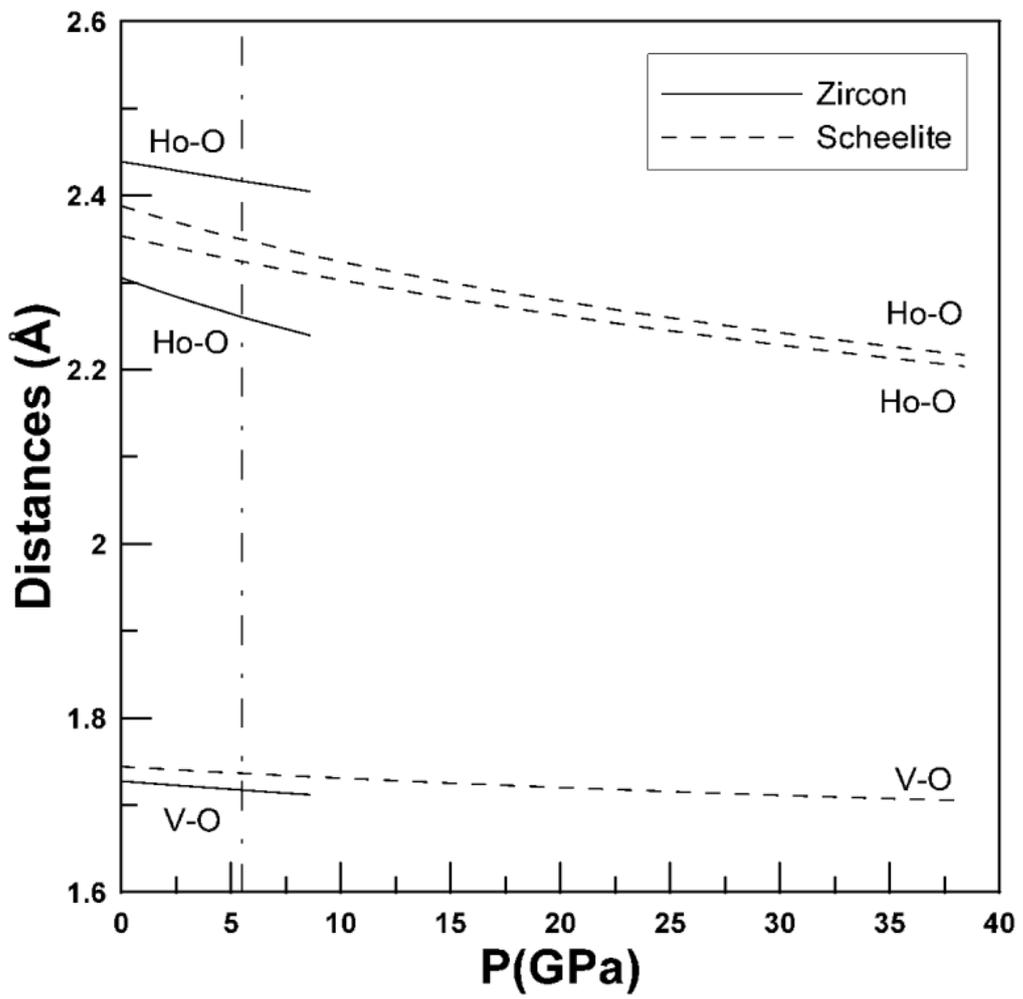